\documentclass[twocolumn,preprintnumbers,superscriptaddress,amsmath,amssymb]{revtex4}
\usepackage{graphicx}
\usepackage{dcolumn}
\usepackage{bm}
\usepackage{amssymb}
\usepackage{amssymb}
\usepackage{verbatim}
\usepackage{color}
\def\>{\rangle}

\usepackage{color}
\usepackage{graphicx}
\usepackage{citesort}
\usepackage{color}
\usepackage{subfigure}
\usepackage{amsmath,amssymb}
\usepackage{upgreek}

\graphicspath{{figs/},}

\begin{document}

\title{Brillouin light storage for 100 pulse widths}

\author{Birgit Stiller}
\email{birgit.stiller@mpl.mpg.de}
\affiliation{Max-Planck-Institute for the Science of Light, Staudtstr. 2, 91058 Erlangen, Germany}
\affiliation{Institute of Photonics and Optical Science (IPOS), School of Physics, University of Sydney, Sydney, 2006 NSW, Australia}
\affiliation{The University of Sydney Nano Institute (Sydney Nano), University of Sydney, Sydney, 2006 NSW, Australia}
\author{Kevin Jaksch}
\affiliation{Max-Planck-Institute for the Science of Light, Staudtstr. 2, 91058 Erlangen, Germany}
\affiliation{Institute of Photonics and Optical Science (IPOS), School of Physics, University of Sydney, Sydney, 2006 NSW, Australia}
\affiliation{The University of Sydney Nano Institute (Sydney Nano), University of Sydney, Sydney, 2006 NSW, Australia}
\author{Johannes Piotrowski}
\affiliation{ETH Zürich, Hönggerbergring 64, 8093 Zürich, Switzerland}
\affiliation{MQ Photonics Research Centre, School of Mathematical and Physical Sciences, Macquarie University, NSW 2109, Australia}
\author{Moritz Merklein}
\affiliation{Institute of Photonics and Optical Science (IPOS), School of Physics, University of Sydney, Sydney, 2006 NSW, Australia}
\affiliation{The University of Sydney Nano Institute (Sydney Nano), University of Sydney, Sydney, 2006 NSW, Australia}
\author{Miko\l{}aj K. Schmidt}
\affiliation{MQ Photonics Research Centre, School of Mathematical and Physical Sciences, Macquarie University, NSW 2109, Australia}
\author{Khu Vu}
\affiliation{Laser Physics Centre, RSPE, Australian National University, Canberra, ACT 0200, Australia}
\author{Pan Ma}
\affiliation{Laser Physics Centre, RSPE, Australian National University, Canberra, ACT 0200, Australia}
\author{Stephen Madden}
\affiliation{Laser Physics Centre, RSPE, Australian National University, Canberra, ACT 0200, Australia}
\author{Michael J. Steel}
\affiliation{MQ Photonics Research Centre, School of Mathematical and Physical Sciences, Macquarie University, NSW 2109, Australia}
\author{Christopher G. Poulton}
\affiliation{School of Mathematical and Physical Sciences, University of Technology Sydney, NSW 2007, Australia}
\author{Benjamin J. Eggleton}
\affiliation{Institute of Photonics and Optical Science (IPOS), School of Physics, University of Sydney, Sydney, 2006 NSW, Australia}
\affiliation{The University of Sydney Nano Institute (Sydney Nano), University of Sydney, Sydney, 2006 NSW, Australia}

\begin{abstract}
Signal processing based on stimulated Brillouin scattering (SBS) is limited by the narrow linewidth of the optoacoustic response, which confines many Brillouin applications to
continuous wave signals or optical pulses longer than several nanoseconds. In this work, we experimentally demonstrate Brillouin interactions at the 150\,ps time scale and a delay for a record 15\,ns which corresponds to a delay of 100 pulse widths. This breakthrough experimental result was enabled by the high local gain of the chalcogenide waveguides as the optoacoustic interaction length reduces with pulse width. We successfully transfer 150ps-long pulses to traveling acoustic waves within a Brillouin-based memory setup. The information encoded in the optical pulses is stored for 15\,ns in the acoustic field. We show the retrieval of eight amplitude levels, multiple consecutive pulses and low distortion in pulse shape. The extension of Brillouin-based storage to the ultra-short pulse regime is an important step for the realisation of practical Brillouin-based delay lines and other optical processing applications.

\end{abstract}

\maketitle

\section{Introduction}

Nonlinear scattering processes such as stimulated Rayleigh, Raman and Brillouin scattering can be distinguished in the optical domain through their different frequency shifts and spectral linewidths. These parameters are characteristic of the light-matter interaction with either macroscopic or molecular vibrations~\cite{Boyd2003,Merklein2022}. Due to its narrow gain linewidth of several tens of MHz, stimulated Brillouin scattering (SBS) is particularly interesting for applications such as optical and microwave filters~\cite{Tanemura2002,Marpaung2015,Choudhary2016,Sancho2010,Vidal2007}, signal processing ~\cite{Santagiustina2013,Shin2015,Song2009,Antman2012,Eggleton2019} and  narrow-linewidth lasers~\cite{Hill1976,Kabakova2013,Gundavarapu2019,Chauhan2021}. On the one hand, the narrow linewidth would suggest that only long pulses can be delayed; the time-bandwidth product then limits the fractional delay. For example SBS-based slow-light schemes are usually  limited to a delay of one or two pulse widths~\cite{Thevenaz2008}. Cascading the SBS slow-light process permits  small improvements in the delay but at the cost of added system complexity~\cite{Song2005}. To overcome these limitations, another concept that uses SBS for delaying optical signals can be used: that of ``Brillouin-based memory'' (BBM)~\cite{Zhu2007,Merklein2017}, which temporarily stores information encoded in light signals as traveling acoustic waves. In these experiments, although there is no slowing of the optical pulse itself, the information can be held 
in a short distance for comparatively long lengths of time, because the processes of acoustic propagation and loss are five orders of magnitude slower than for optical waves.
This concept has been demonstrated in highly nonlinear fibers and integrated photonic circuits~\cite{Zhu2007,Merklein2017} for nanosecond pulses with a storage time of 10 nanoseconds, limited by the acoustic lifetime. The storage can operate at room temperature and can simultaneously store multiple signals that are
closely spaced in frequency without significant cross-talk~\cite{Stiller2019}. In Ref.~\cite{Merklein2017}, the ability to store and retrieve coherent informtation were  demonstrated. Further work showed the non-reciprocity and cascading of the process~\cite{Merklein2020,Stiller2018d}, as well as dynamic reinforcement of the acoustic waves~\cite{Stiller2020}. However, so far experiments in  BBM (as well as in many other Brillouin phenomena), have been limited to nanosecond-long optical pulses~\cite{Shapiro1967}.

\begin{figure*}[t]
  \centering
  \includegraphics[width=1\textwidth]{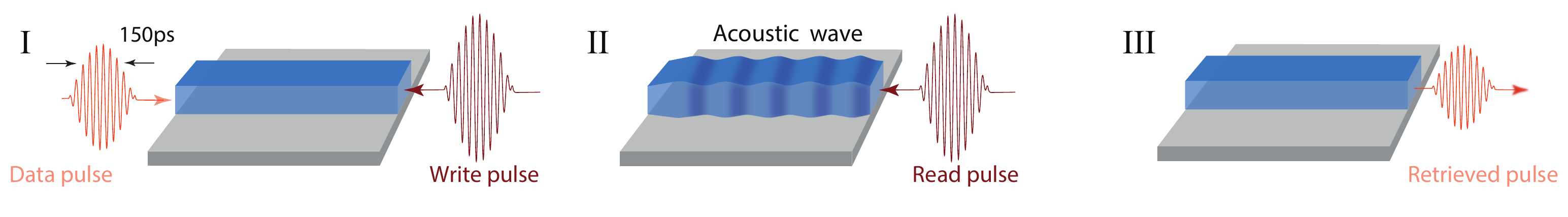}
\caption{Principle of the Brillouin-based memory: I. an optical data pulse of width 150~ps is depleted by a strong counter-propagating ``write'' pulse, storing the data pulse as an acoustic excitation (II). Retrieval process: a read pulse depletes the acoustic wave, converting the data pulse back to the optical domain (III).}
\label{fig1}
\end{figure*}

In this work, we experimentally demonstrate the optoacoustic storage of pulses as short as 150\,ps for up to 15\,ns corresponding to 100 pulsewidths, using the BBM technique. 
We achieve storage of pulses with a linewidth that exceeds the Brillouin linewidth by two orders of magnitude, from 30\,MHz to about 3\,GHz. 
The giant gain of highly-nonlinear chalcogenide waveguides enables highly efficient storage of pulses within cm-scale
waveguides, so that the pulse shapes are not significantly affected by dispersion; 
we store and retrieve eight different power levels, while maintaining pulse width and shape. For pulses with a pulse duration shorter than one nanosecond, we observe a complex behaviour of the read-out efficiency oscillating as a function of the delay time. 
We offer two possible mechanisms for this observation: i) the reach of a transient regime for pulses shorter than 1\,ns that is not well predicted by the usual approximations  made in the coupled mode equations for stimulated Brillouin interactions, or ii) the interference, mediated by the bandwidth of the optical pulse, of an additional acoustic mode. We discuss the scale of these mechanisms and the experimental regimes where they become important.

\section{Experimental results}

To demonstrate the Brillouin interaction of optical pulses shorter than 1\,ns, we use a Brillouin-based light storage setup, as shown in Fig.~\ref{fig1}. 
A ``data'' pulse with central frequency $\omega_d/(2\pi)$ and a Full-Width at Half-Maximum (FWHM) pulse duration $\tau$ enters the Brillouin-active optical waveguide from one side. A counter-propagating ``write'' pulse, at the ``control'' frequency $\omega_c/(2\pi)$ down-shifted by the Brillouin frequency shift of $\Omega/(2\pi) = 7.7 \, \textrm{GHz}$, depletes the data pulse and excites a coherent acoustic wave in the optical waveguide. By sending in a ``read'' pulse at the control frequency $\omega_c/(2\pi)$, the acoustic wave is depleted and a retrieved optical data pulse propagates onwards in the original direction. (Note that here we avoid the conventional nomenclature of ``pump'' for the upper frequency and ``Stokes'' for the lower frequency laser pulses since in Brillouin storage experiments, the Stokes field has the higher intensity).

\begin{figure}[b]
	\centering
	
	\includegraphics[width=0.5\textwidth]{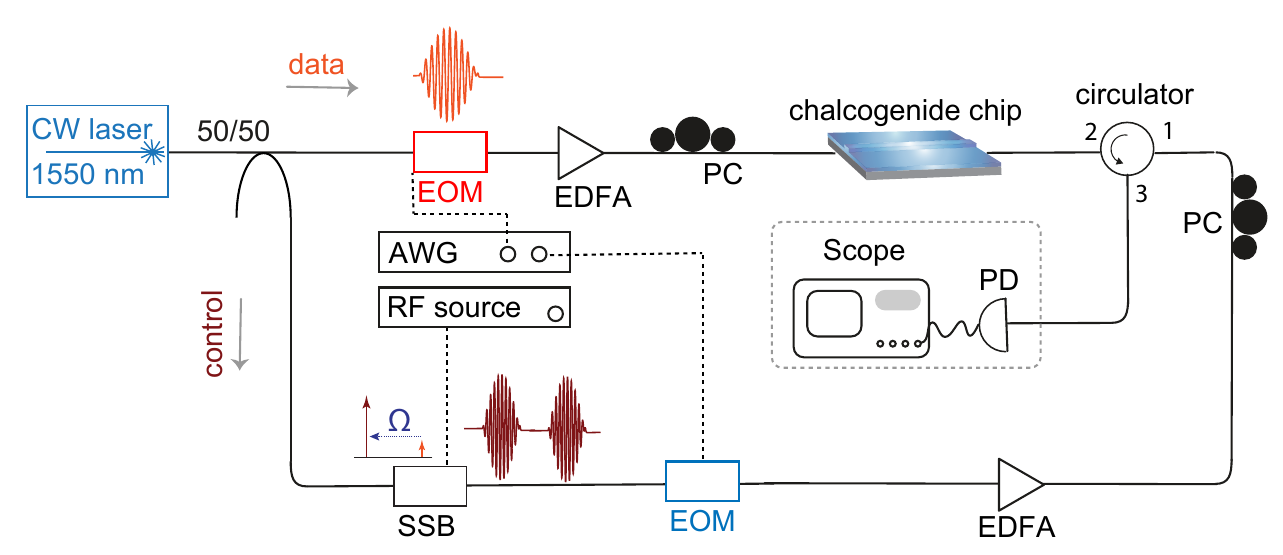} 
	
	\caption{Brillouin-based memory setup. Components in the setup: EDFA, erbium-doped fiber amplifier; PC, polarization controler; SSB, single-sideband modulator; EOM, electro-optic modulator; PD, photo diode; AWG, arbitrary waveform generator; RF, radio frequency. 
	\label{fig2}}
\end{figure}

The  experimental setup is depicted in Fig.~\ref{fig2}. The output of a CW narrow-linewidth distributed feedback laser at 1550\,nm is split into two arms: the data propagation part and the control pulses part. Data pulses with a pulse length down to $\tau=150$\,ps are generated by an electro-optic intensity modulator and a high-speed arbitrary waveform generator. The data pulses are amplified and the polarization adjusted for maximum transmission, before entering the photonic chip from one side with peak power $\sim 50$\,mW.
The As$_2$S$_3$ chalcogenide glass chip (which serves as the storage medium) contains a 17~cm-long small-footprint spiral waveguide with a cross-section of 2.2\,$\upmu$m $\times$ 0.850~$\upmu$m. The CW light on the control arm is frequency down-shifted by the Brillouin frequency shift $\Omega/(2\pi)$ and the respective write and read pulses are imposed on the CW light using an intensity modulator and a second channel of the arbitrary waveform generator. After amplification, the control pulses (peak power $\sim 20$\,W, pulse width $400$\,ps) enter the photonic waveguide from the opposite side (via port 1 of the circulator). The retrieved data pulses exit at port 3 of the circulator before being  filtered with a narrow-linewidth tunable filter and recorded in the time domain with an oscilloscope.

\begin{figure}[b]
  \centering
  \includegraphics[width=0.5\textwidth]{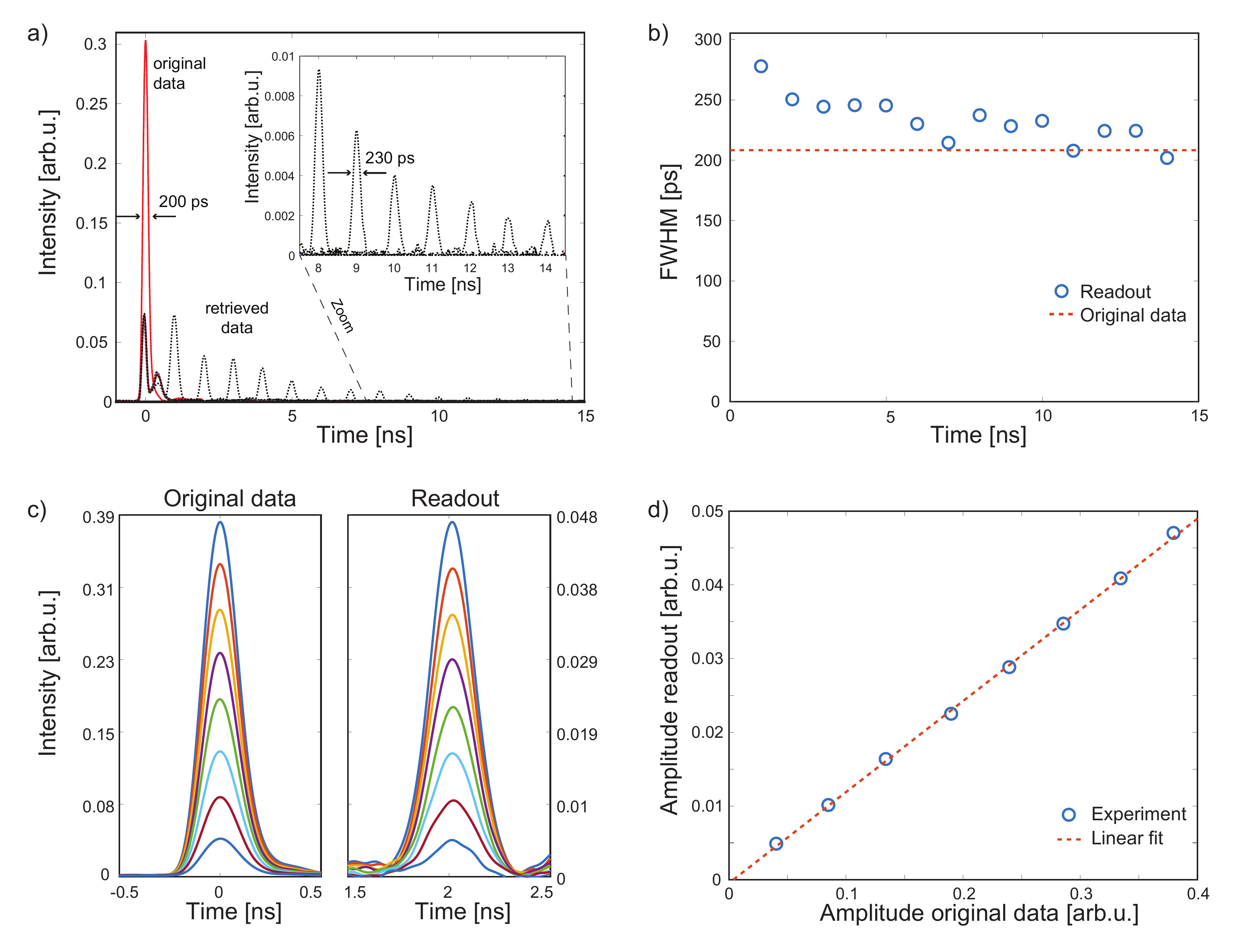}
\caption{Experimental results. a) Tunable storage of a 200~ps-long data pulse for up to 14\,ns. The retrieved data is shown in black dotted lines. (The black signal shown with zero delay is the detected residual power of the data pulse that is not converted to the acoustic wave.) Inset: zoom-in from 8 to 14\,ns. b) Full width at half maximum of the retrieved pulses at different storage times, average value: 230\,ps. c) Read-out of different intensity levels after 2\,ns. d) Linear dependence of input to output amplitude.}
\label{fig3}
\end{figure}

The first experimental results show the storage of 200~ps-long pulses (Fig.~~\ref{fig3}a). The data pulse can be retrieved from the acoustic wave after a storage time which is determined by the time difference between the write and read pulses. We achieved up to 14\,ns delay while maintaining the pulse shape. The pulse width of the retrieved pulses for different storage times is shown in Fig.~\ref{fig3}b, revealing an average pulse width of 230\,ps, only slightly increased compared to the original data pulse. This indicates that our system preserves the pulse linewidth, while the storage time is still limited only by the acoustic lifetime. To further increase the capacity of the BBM, we encode eight different intensity levels on the original data pulse (Fig.~3c, left) which corresponds to 3 bits of information. We store and retrieve the different intensity levels after a storage time of 2\,ns.  As shown in Fig.~3c (right panel), the intensity levels are almost perfectly retrieved, confirming that multiple bits of information can be stored (Fig.~3d).

\begin{figure}[b]
  \centering
  \includegraphics[width=0.45\textwidth]{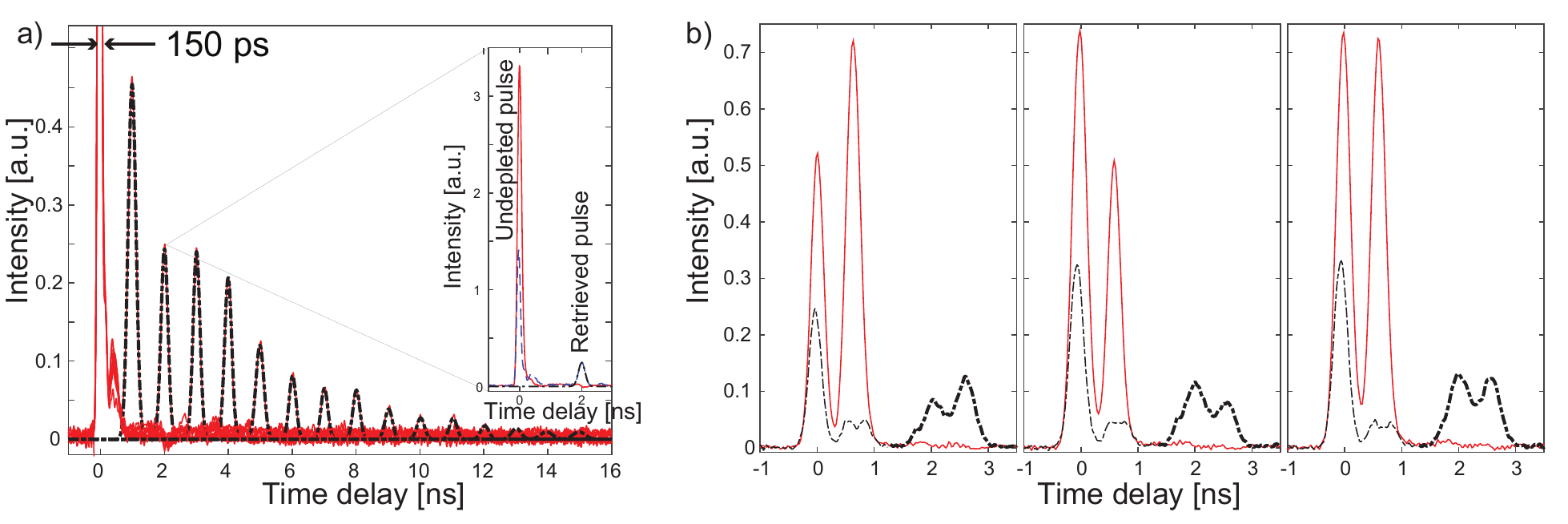}
\caption{Experimental results. A 150\,ps long original data pulse (red) is stored for up to 15\,ns (retrieved signals shown in black).  The inset shows the undepleted pulse compared to the read-out pulse after 2\,ns. b) Two consecutive pulses with a pulse width of 280\,ps and different amplitude levels are stored and retrieved after 2\,ns. The pulse shape and amplitude levels for both optical pulses is retrieved. The detected signal with zero delay is due to the unconverted fraction of the input signal. }
\label{fig4}
\end{figure}

In Fig.~\ref{fig4}a, the same measurement is shown for 150~ps-long pulses, where the data can be retrieved up to 15\,ns after the original pulse. This is a delay of 100 pulse widths,
which is (to the best of our knowledge) a record for Brillouin-based memory. Within this measurement series, we also encoded two amplitude levels on two consecutive short pulses to demonstrate the reliable retrieval of the respective levels and the maintenance of the bandwidth (Fig.~\ref{fig4}b).

In the analysis of both measurements, we observed an anomalous aspect of the  decay of the retrieved data pulses as compared to the normally observed exponential decay (corresponding to the exponential temporal decay of the acoustic wave). In Fig.~\ref{fig3}a there is a smaller intensity than expected at 2\,ns delay time. A similar behaviour can be observed in Fig.~\ref{fig4}a at delays of 2\,ns and 6\,ns. To determine that this was a genuine effect, we performed a more in-depth investigation of this oscillating read-out and  studied the behaviour of the delay and retrieval of pulses with  durations of 280\,ps, 440\,ps, 760\,ps and 1\,ns (see Fig.~\ref{fig5}). In Fig.~\ref{fig5}a, we observe a decay curve very close to exponential,  but as the original data pulse becomes shorter, the departure from simple exponential decay becomes increasingly marked. For a pulse width of 280\,ps and small steps in delay time, the minima and maxima in the read-out efficiency are very pronounced.

\section{Discussion and theoretical considerations}

\begin{figure}[t]
  \centering
  \includegraphics[width=0.5\textwidth]{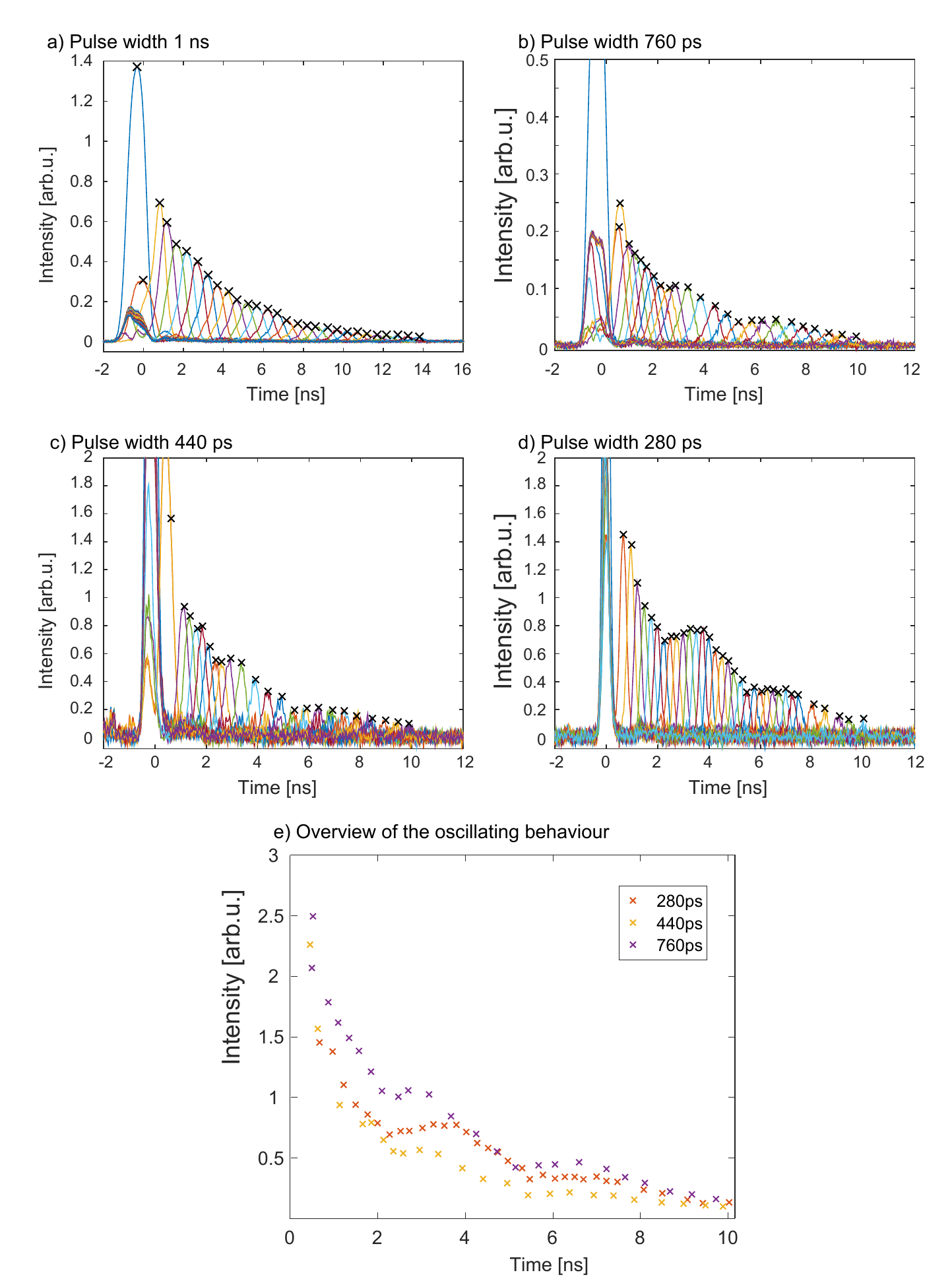}
\caption{Experimental results. For comparison, the tunable storage of optical pulses with different pulse duration is shown: a) 1\,ns, b) 760\,ps, c) 440\,ps and d) 280\,ps. e) Overview on the oscillating behaviour of the read-out amplitude: minima at delays of approximately 2.2\,ns and 5.5\,ns and 9\,ns can be observed.}
\label{fig5}
\end{figure}

There are different possible mechanisms for the oscillating behaviour observed in Fig.~\ref{fig5}. The first is that both the rotating-wave (RWA) and slowly-varying envelope approximations (SVEA) used for modeling SBS have their limitations~\cite{Chamorro-Posada2021}; this issue has been discussed in~\cite{Piotrowski2021} and appears when the acoustic lifetime becomes comparable to the interaction times of sub-nanosecond pulses. For short
pulses, the acoustic envelope $b(z,t)$ satisfies 
\begin{equation}\label{eq:pulseacequation}
     \partial_z b + \frac{i}{2\Omega v_\mathrm{o}}\partial_t^2 b +  \left(\frac{1}{v_\mathrm{o}}+\frac{i\alpha}{2\Omega}\right)\partial_t  b   + \frac{\alpha}{2} b 
    = -iQ \Omega a_{\mathrm{d}}a_{\mathrm{c}}^*\;,
\end{equation}
where $v_\mathrm{o}$ is the  acoustic group velocity, and the spatial (power) damping rate $\alpha=2\Gamma/v_\mathrm{o}$, where $1/\Gamma$ is the acoustic lifetime of the material.
The second term accounts for the non-phase-matched acoustic wave behaviour that arises in the short pulse regime driven by the data $a_{\mathrm{d}}(z,t)$ and control $a_{\mathrm{c}}(z,t)$ pulse envelopes. We use units such that the quantities $|a_{\mathrm{d}}|^2$, $|a_{\mathrm{c}}|^2$ and $|b|^2$ all represent the power (in Watts) carried in each field. Finally,  the nonlinear coupling coefficient $Q$, which is related to the steady-state gain $g_0$, as measured in m$^{-1}$W$^{-1}$, 
by $g_0 = 2 \omega_d \Omega |Q|^2/\alpha$~\cite{wolff2021}.

The presence of oscillations  depends on the contribution of the higher order terms in equation~(\ref{eq:pulseacequation}), which become relevant below $(2\pi)/\Omega \approx 130\,\mathrm{ps}$ pulse lengths. In the case of even shorter pulses than reported here, the acoustic envelope will oscillate at $f_\mathrm{B} = 7.7\,\mathrm{GHz}$. According to~\cite{Piotrowski2021}, the current experiments lie at the limit of the resolution where the oscillation can be observed. However, due to a possible frequency chirp in the experimental pulses, the regime of an oscillating read-out efficiency might have been reached.

\begin{figure}  \centering
  \includegraphics[width=\linewidth]{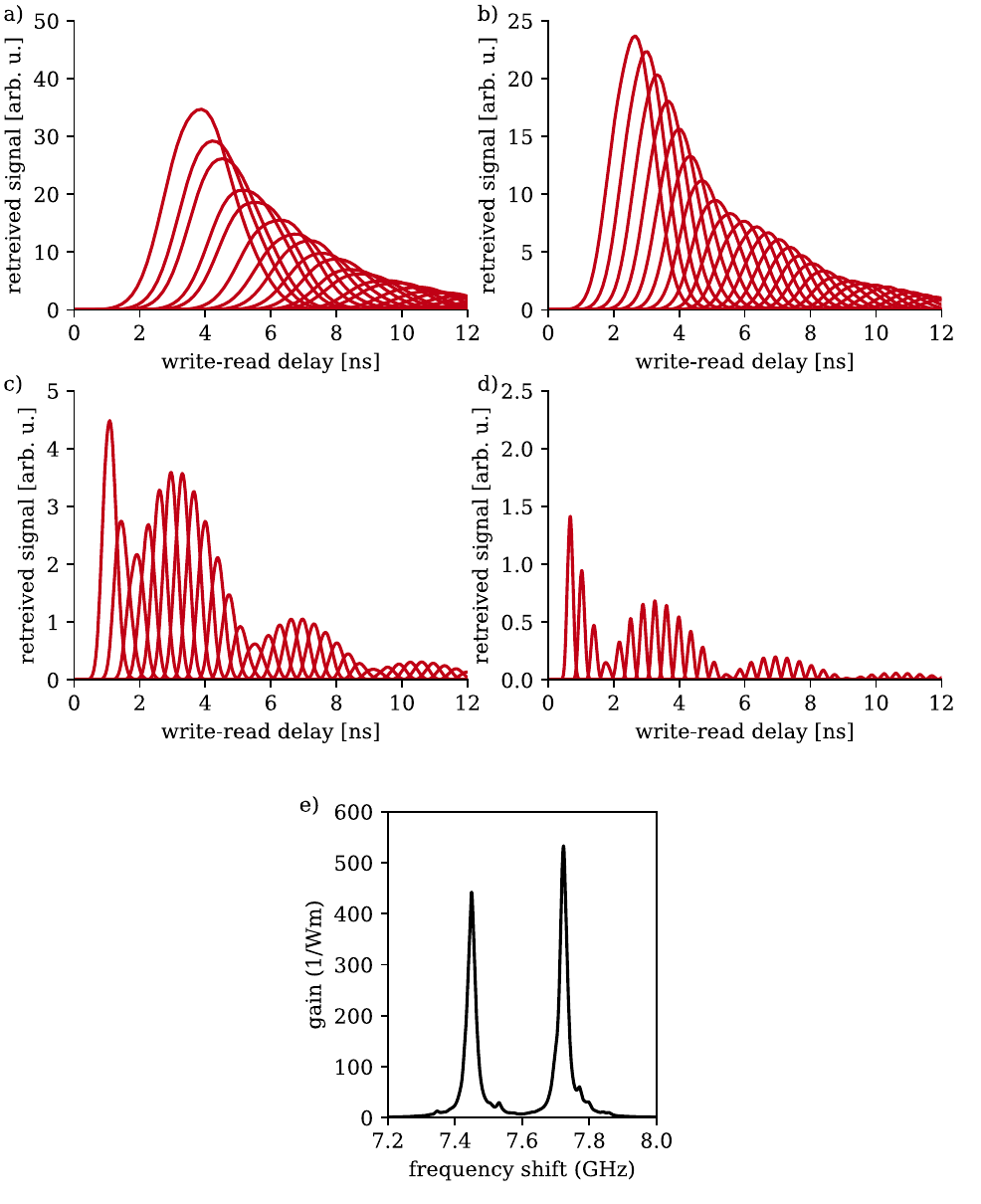}
\caption{Simulated readout signal of two acoustic modes after storage process of a) 5\,ns, b) 3\,ns, c) 1\,ns, d) 0.44\,ns input signal pulses with the unconverted input removed. e) NumBAT simulation of Brillouin gain from an As$_2$S$_3$ slot waveguide ($2200\,\textrm{nm}\times850\,\textrm{nm}$ slot embedded in $4250\,\textrm{nm}$ high slab) shows two adjacent, supported acoustic modes with similar gain.}
\label{fig7}
\end{figure}

A second possible reason for the oscillating behaviour at longer timescales is the coupling of the optical fields through a second acoustic mode. With the broader spectrum of shorter pulses, two acoustic modes of similar frequencies can be driven simultaneously and can interfere. In Fig.~\ref{fig7} we show optoacoustic calculations for our As$_2$S$_3$ waveguide with the open source NumBAT~\cite{Sturmberg2019} software and find a second acoustic mode with a frequency shift $\Delta/( 2\pi) = (\Omega_1-\Omega_2)/( 2\pi)  = (7.72- 7.45)\,\textrm{GHz}=  270\,\textrm{MHz}$ from the targeted Brillouin mode.

Simulations of the acoustic modes in this waveguide reveal that there are in fact several nearby acoustic modes that can, in principle, couple to the optical field. For the waveguide used in our experiments, these modes undergo an anti-crossing very near to the point where the optical and acoustic modes are phase-matched. This anti-crossing can cause significant variation in the acoustic frequencies in response to small changes in waveguide geometry, such as rib thickness and height. For all following simulations we assume an ``ideal'' geometry, in which the secondary mode lies at 270MHz. This results in Brillouin frequencies of $\Omega_1/(2\pi)=$7.72\,GHz and $\Omega_2/(2\pi)=$7.45\,GHz, as well as gain values of $g_{0,1}=550\,(\textrm{Wm})^{-1}$ and $g_{0,2}=440\,(\textrm{Wm})^{-1}$.

Experimental verification with a pump-probe setup did not reveal the existence of a strong second acoustic mode at 7.45\,GHz. However, there is the possibility of a very inefficiently coupled mode that was not detectable with a pump-probe setup that still influences the writing and reading process in the Brillouin-based memory setup. Note that the efficiency of the gain strongly depends on the overlap of the optical and the acoustic modes, which can lead to an experimental outcome where only one of the modes can be observed through pump-probe measurements. 

On incorporating a second acoustic mode at frequency $\Omega_2$ so that the displacement field takes the form

\begin{equation*}\label{eq:2mode}
    \mathbf { U } ( \mathbf { r } , t ) = \tilde { \mathbf { u } }_1 ( x , y ) e^{i(  q z -  \Omega_1 t )} b_1 ( z , t ) +
\end{equation*}
\begin{equation}\label{eq:2mode}
   + \tilde { \mathbf { u } }_2 ( x , y ) e^{i(  q z -  \Omega_2 t )} b_2 ( z , t ) + \mathrm { c . c . }\;, 
\end{equation}
and then applying the RWA and SVEA in the usual fashion~\cite{Wolff2015} we can obtain the following coupled mode equations for the two optical waves and  two acoustic waves:

\begin{align}\label{cm1}
    \frac{1}{v_\mathrm{o_1}} \partial_t b_1 
    +\partial_z b_1 +
    \alpha_1 b_1&= 
    -i\Omega_1 Q_\mathrm{1} a_{\mathrm{d}}a_{\mathrm{c}}^*  \
    \\\label{cm2}
    \frac{1}{v_\mathrm{o_2}}\partial_tb_2 
    +\partial_z b_2 +
    \alpha_2b_2&
    = -i\Omega_2 Q_\mathrm{2} a_{\mathrm{d}}a_{\mathrm{c}}^* e^{i t\Delta}\\
    \partial_{z} a_{\mathrm{d}}+\frac{1}{v_{\mathrm{d}}} \partial_{t} a_{\mathrm{d}} &=-i \omega_{\mathrm{d}} \left( Q_{\mathrm{1}} a_{\mathrm{c}} b_1 + e^{-i t\Delta}Q_{2} a_{\mathrm{c}} b_2\right)
,
\\ \partial_{z} a_{\mathrm{c}}-\frac{1}{v_{\mathrm{c}}} \partial_{t} a_{\mathrm{c}} &=-i \omega_{\mathrm{c}}  \left(Q_{\mathrm{1}} a_{\mathrm{d}} b_1^* + e^{i t\Delta}Q_{\mathrm{2}} a_{\mathrm{d}} b_2^*\right)\;.
\label{cm4}\end{align}
For the two acoustic modes with similar mode profile we can assume ${v_\mathrm{o_1}} = {v_\mathrm{o_2}} \approx 0$ and $\alpha_1 = \alpha_2$, and we take  $v_d=v_c>0$. In contrast to the usual formalism,  Eq.~(\ref{cm2}) is added to capture the second acoustic mode with an oscillating driving term from the mismatched Brillouin shift. Corresponding oscillatory factors appear in the new optical driving terms.
In Fig.~\ref{fig7} we simulate~\cite{Piotrowski2021} equations (\ref{cm1})-(\ref{cm4}) with 3\,ns exponential decay factor for 50\,mW data and 20\,W control peak powers for $\tau=$5\,ns, 3\,ns, 1\,ns and 0.44\,ns pulses. Wave velocities in the chalcogenide waveguide (n = 2.37) taken as $v_{o1} = v_{o2} = 2595\textrm{m}\textrm{s}^{-1}$ and $v_{d} = v_{c} \approx 1.26\times10^8\textrm{m}\textrm{s}^{-1}$. The retrieved signal after the write-read process picks up the modulation at $\Delta/(2\pi) =  270\,\textrm{MHz}$, as well as the exponential decay from the acoustic lifetime. For short pulses (Fig.~\ref{fig7}d), this qualitatively reproduces the optical readout seen in Fig.~\ref{fig5}d. 

We also observe the decrease in visibility of the modulation for longer pulses. The modulations with frequency $\Delta$ are averaged over as $\tau$ approaches $2\pi/\Delta \approx 3.7\,\textrm{ns}$. As can be seen in Fig.~\ref{fig7}, the modulation begins to be discernible for pulses with lengths less than 3~ns and is clearly visible for sub-ns pulses. While this is for longer pulse lengths than the observed modulation in the experimental data (Fig.~\ref{fig5}), which becomes discernible only in the sub-ns range, it is within the same order of magnitude and so remains a plausible mechanism for the experimental observations. These simulations demonstrate that coupling to existing secondary acoustic modes becomes increasingly important for shorter pulse lengths.

\begin{table*}
\begin{tabular}{c|c}
Quantity &  Value \\ \hline
     $v_c,v_d$ &  $c/2.37$ \\
     $\alpha$ &  $128452 \textrm{m}^{-1}$ (3~ns decay time at $2595\textrm{m}\textrm{s}^{-1}$) \\
     $P_c$ & 20 W\\
     $P_d$ & 1 mW \\
     $\tau_c$ & 400 ps\\
     $\tau_d$ & 280-1000 ps \\
     $Q_1$ &  $7.75\times10^{-10} \textrm{ s}\textrm{m}^{-1}\textrm{W}^{-1/2}$ ($g_0 = 2 \omega_d \Omega |Q|^2/\alpha$, with $g = 550\textrm{ W}^{-1}\textrm{m}^{-1}$)\\
     $Q_2$ &  $7.05\times10^{-10} \textrm{ s}\textrm{m}^{-1}\textrm{W}^{-1/2}$ 
\end{tabular}
\label{tab:parameters1}
\caption{Parameters used in the numerical model of equations (\ref{cm1}--\ref{cm4}). Note that the quantities $v_{o_1}$ and $v_{o_2}$ appearing in these equations have been assumed to be zero.}
\end{table*}
 
\section{Conclusion}

We have demonstrated stimulated Brillouin interactions with short pulses down to the 150\,ps regime. With the concept of Brillouin-based memory we showed that the limitations of the narrow Brillouin linewidth can be overcome by two orders of magnitude. Data pulses down to the 150\,ps regime can be retrieved for up to 100 pulse widths while maintaining the pulse bandwidth, pulse shape and amplitude encoded information. This result together with the previously shown coherence~\cite{Merklein2017} and frequency preserving nature of the BBM~\cite{Stiller2019} makes this approach suitable for advanced coherent communication schemes. Finally, we demonstrated a thorough measurement of the read-out intensity dependent on the storage time for different pulse widths which revealed an oscillating behaviour. We demonstrated with simulations that this can be explained by the coupling to a second acoustic mode whereby the interference depth of the oscillation is mediated by the bandwidth of the optical pulses. There is also the possibility that the usual coupled mode equations using the slowly varying amplitude approximation are not valid anymore in the short pulse regime below 1\,ns which would equally lead to an oscillating behaviour with fast oscillations \cite{Piotrowski2021}.


\section*{Acknowledgements}
This work was sponsored by the Australian Research Council (ARC) Laureate Fellowship (FL120100029) and the Centre of Excellence program (CUDOS CE110001010), as well as through Discovery Projects DP200101893 and DP220101431. We acknowledge the support of the ANFF ACT and funding from the Independant Max Planck Research Group Scheme.

\section*{Data availability}
Source data are available for this paper. All other data that support the plots within this paper and other findings of this study are available from the corresponding author upon reasonable request.

\section*{Author Contributions Statement}

B.S., K.J., and M.M. conceived the idea. B.S., K. J., and M. M. performed the experiments and analyzed the data. J.P., M.K.S., C.G.P. and M.J.K. developed the theory. K.V., P.M. and S.M. fabricated the sample. B.S., J.P., M.K.S., C.G.P., M.J.K., M.M. and B.J.E. wrote the manuscript with input from all authors.

\section*{Competing Interests Statement}
The authors declare no competing interests.

\section{References}


\bibliographystyle{iopart-num}
\bibliography{library_updated3.bib,extras}

\providecommand{\newblock}{}
\begin{thebibliography}{10}
\expandafter\ifx\csname url\endcsname\relax
  \def\url#1{{\tt #1}}\fi
\expandafter\ifx\csname urlprefix\endcsname\relax\def\urlprefix{URL }\fi
\providecommand{\eprint}[2][]{\url{#2}}

\bibitem{Boyd2003}
Boyd R~W 2003 {\em {Nonlinear Optics}\/} (Acad. Press) ISBN 0121216829

\bibitem{Merklein2022}
Merklein M, Kabakova I~V, Zarifi A and Eggleton B~J 2022 {\em Applied Physics
  Reviews\/} {\bf 9} 041306 ISSN 1931-9401

\bibitem{Tanemura2002}
Tanemura T, Takushima Y and Kikuchi K 2002 {\em Optics Letters\/} {\bf 27} ISSN
  0146-9592

\bibitem{Marpaung2015}
Marpaung D, Morrison B, Pagani M, Pant R, Choi D~Y, Luther-Davies B, Madden S~J
  and Eggleton B~J 2015 {\em Optica\/} {\bf 2} 76 ISSN 2334-2536
  \urlprefix\url{http://www.opticsinfobase.org/abstract.cfm?URI=optica-2-2-76}

\bibitem{Choudhary2016}
Choudhary A, Aryanfar I, Shahnia S, Morrison B, Vu K, Madden S, Luther-Davies
  B, Marpaung D and Eggleton B~J 2016 {\em Optics Letters\/} {\bf 41} 436--439

\bibitem{Sancho2010}
Sancho J, Chin S, Sagues M, Loayssa A, Lloret J, Gasulla I, Sales S, Thevenaz L
  and Capmany J 2010 {\em IEEE Photonics Technology Letters\/} {\bf 22}
  1753--1755 ISSN 10411135

\bibitem{Vidal2007}
Vidal B, Piqueras M~a and Mart{\'{i}} J 2007 {\em Optics Letters\/} {\bf 32}
  23--25 ISSN 0146-9592

\bibitem{Santagiustina2013}
Santagiustina M, Chin S, Primerov N, Ursini L and Th{\'{e}}venaz L 2013 {\em
  Scientific Reports\/} {\bf 3} ISSN 2045-2322

\bibitem{Shin2015}
Shin H, Cox J~A, Jarecki R, Starbuck A, Wang Z and Rakich P~T 2015 {\em Nature
  Communications\/} {\bf 6} 6427 ISSN 2041-1723
  \urlprefix\url{http://www.ncbi.nlm.nih.gov/pubmed/25740405}

\bibitem{Song2009}
Song K~Y, Lee K and Lee S~B 2009 {\em Optics Express\/} {\bf 17} 10344--9 ISSN
  1094-4087 \urlprefix\url{http://www.ncbi.nlm.nih.gov/pubmed/19506688}

\bibitem{Antman2012}
Antman Y, Levanon N and Zadok A 2012 {\em Optics Letters\/} {\bf 37} 5259--61
  ISSN 1539-4794 \urlprefix\url{http://www.ncbi.nlm.nih.gov/pubmed/23258071}

\bibitem{Eggleton2019}
Eggleton B~J, Poulton C~G, Rakich P~T, Steel M and Bahl G 2019 {\em Nat.
  Photonics\/} {\bf 13} 664--677

\bibitem{Hill1976}
Hill K~O, Kawasaki B~S and Johnson D~C 1976 {\em Applied Physics Letters\/}
  {\bf 28} 608 ISSN 00036951
  \urlprefix\url{http://link.aip.org/link/APPLAB/v28/i10/p608/s1{\&}Agg=doi}

\bibitem{Kabakova2013}
Kabakova I~V, Pant R, Choi D~Y, Debbarma S, Luther-Davies B, Madden S~J and
  Eggleton B~J 2013 {\em Optics Letters\/} {\bf 38} 3208--11 ISSN 1539-4794
  \urlprefix\url{http://www.ncbi.nlm.nih.gov/pubmed/23988915}

\bibitem{Gundavarapu2019}
Gundavarapu S, Brodnik G~M, Puckett M, Huffman T, Bose D, Behunin R, Wu J, Qiu
  T, Pinho C, Chauhan N, Nohava J, Rakich P~T, Nelson K~D, Salit M and
  Blumenthal D~J 2019 {\em Nature Photonics\/} {\bf 13}(1) ISSN 17494893

\bibitem{Chauhan2021}
Chauhan N, Isichenko A, Liu K, Wang J, Zhao Q, Behunin R~O, Rakich P~T, Jayich
  A~M, Fertig C, Hoyt C~W and Blumenthal D~J 2021 {\em Nature Communications\/}
  {\bf 12} 4685 ISSN 2041-1723
  \urlprefix\url{https://doi.org/10.1038/s41467-021-24926-8}

\bibitem{Thevenaz2008}
Th{\'{e}}venaz L 2008 {\em Nature Photonics\/} {\bf 2} 474--481 ISSN 1749-4885
  \urlprefix\url{http://www.nature.com/nphoton/journal/v2/n8/abs/nphoton.2008.147.html
  http://www.nature.com/doifinder/10.1038/nphoton.2008.147}

\bibitem{Song2005}
Song K~Y, Herr{\'{a}}ez M and Th{\'{e}}venaz L 2005 {\em Optics Express\/} {\bf
  13} 82--88 ISSN 1094-4087

\bibitem{Zhu2007}
Zhu Z, Gauthier D~J and Boyd R~W 2007 {\em Science\/} {\bf 318} 1748--50 ISSN
  1095-9203 \urlprefix\url{http://www.ncbi.nlm.nih.gov/pubmed/18079395}

\bibitem{Merklein2017}
Merklein M, Stiller B, Vu K, Madden S~J and Eggleton B~J 2017 {\em Nature
  Communications\/} {\bf 8} 574 ISSN 2041-1723
  \urlprefix\url{http://www.nature.com/articles/s41467-017-00717-y}

\bibitem{Stiller2019}
Stiller B, Merklein M, Vu K, Ma P, Madden S~J, Poulton C~G and Eggleton B~J
  2019 {\em APL Photonics\/} {\bf 4} ISSN 23780967

\bibitem{Merklein2020}
Merklein M, Stiller B, Vu K, Ma P, Madden S~J and Eggleton B~J 2020 {\em
  Nanophotonics\/} ISSN 21928614

\bibitem{Stiller2018d}
Stiller B, Merklein M, Wolff C, Vu K, Ma P, Poulton C~G, Madden S~J and
  Eggleton B~J 2018 {\em Optics Letters\/} {\bf 43}(18) ISSN 0146-9592

\bibitem{Stiller2020}
Stiller B, Merklein M, Wolff C, Vu K, Ma P, Madden S~J and Eggleton B~J 2020
  {\em Optica\/} {\bf 7} ISSN 2334-2536

\bibitem{Shapiro1967}
Shapiro S~L, Giordmaine J~A and Wecht K~W 1967 {\em Physical Review Letters\/}
  {\bf 19}(19) ISSN 00319007

\bibitem{Chamorro-Posada2021}
Chamorro-Posada P and de~la Llera J~B 2021 {\em Optical Fiber Technology\/}
  {\bf 65} 102592 ISSN 1068-5200

\bibitem{Piotrowski2021}
Piotrowski J, Schmidt M~K, Stiller B, Poulton C~G and Steel M~J 2021 {\em
  Optics Letters\/} {\bf 46} ISSN 0146-9592

\bibitem{wolff2021}
Wolff C, Smith M~J~A, Stiller B and Poulton C~G 2021 {\em J. Opt. Soc. Am. B\/}
  {\bf 38} 1243--1269
  \urlprefix\url{http://opg.optica.org/josab/abstract.cfm?URI=josab-38-4-1243}

\bibitem{Sturmberg2019}
Sturmberg B~C~P, Dossou K~B, Smith M~J~A, Morrison B, Poulton C~G and Steel M~J
  2019 {\em J. Lightwave Technol.\/} {\bf 37} 3791--3804
  \urlprefix\url{http://opg.optica.org/jlt/abstract.cfm?URI=jlt-37-15-3791}

\bibitem{Wolff2015}
Wolff C, Steel M~J, Eggleton B~J and Poulton C~G 2015 {\em Physical Review A\/}
  {\bf 92} 013836 ISSN 1050-2947
  \urlprefix\url{http://link.aps.org/doi/10.1103/PhysRevA.92.013836}

\end{thebibliography}

\end{document}